\documentstyle[prl,aps]{revtex}

\newcommand{\be}{\begin{equation}}
\newcommand{\ee}{\end{equation}}
\newcommand{\ber}{\begin{eqnarray}}
\newcommand{\eer}{\end{eqnarray}}

\begin{document}
%\draft
%\tighten
%\widetext
\twocolumn[{
\narrowtext 
\title{Tunneling into the edge of a compressible Quantum Hall state}
\author{ A.~V.~Shytov$^a$, L.~S.~Levitov$^a$, and B.~I.~Halperin$^{b,c}$}
\address{
{\it (a)} 12-112, Massachusetts Institute of Technology, Cambridge MA 02139;
{\it (b)} Physics Department, Harvard University, Cambridge MA 02138;
{\it (c)} Institute for Advanced Study, Princeton NJ 08540}
\maketitle
\mediumtext
\begin{abstract}
   We present a composite fermion theory of tunneling into the edge of a
compressible quantum Hall system. The
    tunnel
conductance is non-ohmic, due to slow relaxation of
electromagnetic and Chern-Simons field disturbances caused by
the tunneling electron. Universal results are obtained in the
limit of a large number of channels involved in the relaxation.
The tunneling exponent is found to be a continuous function of
the filling factor $\nu$, with a a slope that is discontinuous
at $\nu=1/2$ in the limit of vanishing bulk resistivity
$\rho_{xx}$. When $\nu$ corresponds to a principal fractional
quantized Hall state, our results agree with the chiral
Luttinger liquid theories of Wen, and Kane, Fisher and
Polchinski.
   \end{abstract}
}]
 
\narrowtext
\noindent {\it Introduction:}
   The edge of a Quantum Hall (QH) system plays a central role in
charge transport, because the edge states carry Hall
current\cite{edge}. Also, for
odd-denominator Landau level filling factors $\nu$ that
correspond to incompressible quantized Hall states,
the excitations on the edge form a
strongly interacting one-dimensional system, which has drawn a lot of
interest\cite{Wen,Fisher}. The theoretical picture of the QH edge is
based on chiral Luttinger liquid models, involving either one or
several chiral modes which may travel in the same or in opposite
directions.
 
   Another important part of the QH theory is the
fermion-Chern-Simons approach, which can describe compressible QH states
at even-denominator fractions such as $\nu=1/2$, as well as the
incompressible states\cite{HLR,Lopez}.
In this approach, the fractional QH effect
is mapped onto the integer QH problem for new quasiparticles,
composite fermions\cite{Jain}, which interact with  a statistical
Chern-Simons
gauge field such that each fermion carries with it  an even
number $p$ of quanta of the Chern-Simons magnetic flux.
The structure
of the edge can then be obtained from Landau levels for composite
fermions in the average residual magnetic field\cite{CFedge}.
 
The physics of the edge can be probed experimentally by a
tunneling conductivity measurement. Soon after first attempts to
study tunneling between edges of incompressible QH states
in conventional gated 2D structures\cite{Webb}, a new
generation of 2D systems was developed by using the cleaved edge
overgrowth technique\cite{Chang1}. In these structures it is
possible to study tunneling into the edge of a 2D electron gas
from a 3D doped region.
It is believed that the 2D electron system has a sharp edge, and that the
the confining potential is very smooth, with
residual roughness of an atomic scale.
With the advantge of the high
quality of the system, one can explore tunneling into both
incompressible and compressible QH states\cite{Chang2}. It is
found that the tunneling conductivity is non-ohmic, $I\sim
V^{\alpha}$, for $V > 2 \pi k_B T / e$,
where the exponent $\alpha$ is a continuously decreasing
function of the filling factor $\nu$\cite{Chang3}.
 
The Luttinger liquid theories predict a power law conductivity, in
agreement with the experiment\cite{Chang2}. However, such
theories can be constructed only for specific incompressible
densities. Furthermore, in the construction there is no
continuity in the filling factor, because the number of chiral
modes, their propagation direction, and the couplings change
from one incompressible state to the other. Thus, it is of
interest to develop an alternative theory capable of treating
both compressible and incompressible states on a similar footing.
 
In this paper we present a theory of tunneling into the QH edge
based on the composite fermion picture. We find that under
certain conditions the tunneling exponent depends only on the
conductivity and interaction in the bulk, and is insensitive to
the detailed structure of the edge. In this case the main effect
results from the relaxation of electromagnetic disturbance
caused by tunneling electron, in which we include charge and
current densities as well as the Chern-Simons field. The
characteristic times and spatial scales involved in the dynamics
are very large, which makes it possible to express the
leading behavior in terms of measurable electromagnetic
response functions, the longitudinal and Hall conductivities.
Tunneling exponents can then be found as a function of the filling
factor.
 
Let us mention that tunneling density of states in a different
model of a compressible edge was studied recently by Conti and
Vignale\cite{Conti-Vignale}, by Han and
Thouless\cite{Han-Thouless}, and by Han\cite{Han}. These works
focus on the case of electron density profile in which the
density is smoothly decreasing from the bulk value $\nu_{\rm
bulk}$ to zero at the boundary. It is assumed that the width of
the region in which the density is changing is many magnetic
lengths. For such an edge there are many hydrodynamical modes
whose contribution to the tunneling density of states must be
taken into account. The tunneling exponent obtained in this
model is much larger than $1/\nu_{\rm bulk}$.
 
\noindent {\it Main results:\hskip2mm}
In the calculation, we assume that the number of flux quanta
carried by composite fermions is $p=2$ for $1/3<\nu<1$ and $p=4$
for $1/5<\nu<1/3$. Also, for simplicity, we assume that the composite
fermions have "bare" conductivites
$\rho^{(0)}_{xx}$ and
$\rho^{(0)}_{xy}$ which are constants that may depend, for instance, on
the
density, but which are independent of temperature. The measured
resistivities are then $\rho_{xy}=
\rho^{(0)}_{xy}+p h/e^2$ and $\rho_{xx}=\rho^{(0)}_{xx}$.
The theory predicts the power-law $I\sim V^{\alpha}$, with
   \be\label{Alpha}
\alpha=1+
{2e^2 \over\pi h}\Big[\theta_H\rho_{xy}-\theta^{(0)}_H\rho^{(0)}_{xy}\Big] +
{e^2\rho_{xx} \over\pi h}\ln{\kappa_0\sigma_{xx}\over \kappa\sigma^{(0)}_{xx}}
   \ee
where $\theta_H=\tan^{-1}\rho_{xx}/\rho_{xy}$ is the Hall angle,
$\theta^{(0)}_H$ is the corresponding bare Hall angle, $\kappa$ is the
compressibility, and
$\kappa_0= m^\ast/2\pi\hbar^2$ is the bare composite-fermion
compressibility, determined by the effective mass
$m^\ast$, which we treat as a constant.
We have assumed, for simplicity, a short-range repulsion
between the electrons (see Fig.~1).
\input epsfig.sty
%\begin{center}
\begin{figure}
%\figure
\label{fig1}
\epsfig{file= 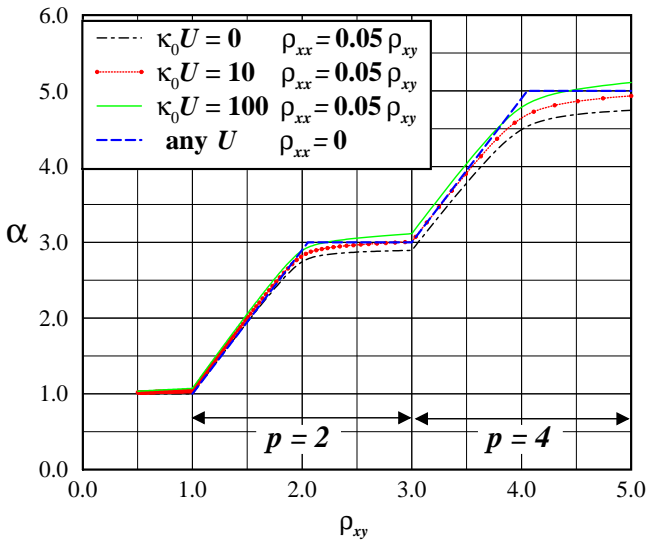, height= 75mm,  angle= 270}
%\epsffile{figure1.eps}
\caption{
The tunneling exponent (\protect{\ref{Alpha}}) is shown as a
function of $\rho_{xy}$ with the composite fermion flux $p=2$
for $1\le\rho_{xy}\le 3$ and $p=4$ for $3\le\rho_{xy}\le5$.
Constant Hall angle is assumed. For $\rho_{xx}=0.05\rho_{xy}$
the exponent is plotted for three values of the model
short-range interaction $U=1/\kappa-1/\kappa_0$. At $\rho_{xx}=0$
the exponent is universal (no $U-$dependence), but at finite
$\rho_{xx}$ it can be either bigger or smaller than the
universal result, depending on the interaction strength
$\kappa_0U$.
}
\end{figure}
%\end{center} 
It is interesting to compare this result with theories of tunneling
into the edge of an incompressible quantized Hall state, where the
edge is described as a
single- or multi-channel chiral Luttinger liquid.
For Jain filling fractions $\nu=n/(n p+1)$
with positive integer $n$ and even $p$, there are $n$ edge modes, all
traveling in the same direction.   In this case, Wen\cite{Wen} found
universally $\alpha=p+1$.
By contrast, for  states such as the Jain fractions with negative $n$,
where the edges have modes going in opposite directions, $\alpha$ may
depend on the form of the interaction.
Nevertheless, Kane, Fisher and Polchinski\cite{Fisher} found in this
case that
if there is sufficient scattering between the channels,
the system scales to a universal
limit, with
$\alpha=p+1-2/|n|$.
 
Our result (\ref{Alpha}), taken at
large Hall angle ($\theta_H=\theta^{(0)}_H=\pi/2$),
agrees with the chiral Luttinger liquid theories discussed
above, when $\nu$ corresponds to a Jain fraction. Specifically, with
  \be
\rho_{xy}=(p+1/n){h\over e^2},\ \rho^{(0)}_{xy}={h\over ne^2},\
\rho_{xx}=\rho^{(0)}_{xx}=0,\
  \ee
substituted in Eq.~(\ref{Alpha}), one gets
$\alpha=1+|p+1/n|-1/|n|$, in agreement with the universal
tunneling exponents predicted by Wen and by Kane et al.
 
However, the theory presented below is
not restricted to the incompressible states. The tunneling
exponent (\ref{Alpha}) is a continuous function of the filling
factor $\nu$. It is interesting, however, that  $\alpha$ has
cusp-like singularities at $\nu=1/2$ and $\nu=1/4$,
in the limit $\rho_{xx}=0$.
To understand this, consider the vicinity of
$\nu=1/2$, where the QH state can be
described as a Fermi liquid of composite fermions carrying two
flux quanta each, and exposed to ``residual'' magnetic field
$\delta B=2-\nu^{-1}$. At $\nu<1/2$ the residual field direction
coincides with total field, and all edge modes propagate in the
same direction\cite{CFedge}. On the other hand, at $\nu>1/2$,
the structure of the edge is qualitatively different, consisting
of modes going in opposite directions. This effect makes
$\nu=1/2$ a singular density. Of course, scattering by disorder
will smear the singularity. However, it is interesting that the
change in the tunneling exponent (\ref{Alpha}) resulting from
finite $\rho_{xx}$ can be either positive or negative, depending
on the value of $\kappa_0 / \kappa $ (see Fig.~1).
 
\noindent {\it Semiclassical Green's function:\hskip2mm}
  We show below that the many-body effect on the tunneling rate
arises mainly from the interaction of tunneling electron with
slow electrodynamical modes. The low-energy QH physics can be
described\cite{HLR} by transforming electrons into composite
fermions interacting via statistical Chern-Simons gauge field
$a_\mu({\bf r}, \, t)$ ($\mu=0,1,2$). At low energies, the
electrodynamical modes can be derived from an effective action
$S_{eff}[a_{\mu}]$ for $a_{\mu}$\cite{KimWen}. To evaluate the
effect of slow fluctuations of $a_{\mu}({\bf r}, \, t)$ on the
electron Green's function, we write it as the average:
   \begin{equation}
\label{start}
G_{{\bf r}\,{\bf r'}}(\tau) =
\frac{{\displaystyle
\int Da_{\mu} \, G_{{\bf r} \, {\bf r'}}(\tau, \, a_{\mu}) \,
e^{-S_{eff}[a_\mu]}} }{{\displaystyle  \int Da_{\mu} \,
e^{-S_{eff}[a_\mu]} } }
  \end{equation}
Here $G_{{\bf r} \,{\bf r'}}(\tau \, a_{\mu})$ is the electron
Green's function for a given configuration of the gauge field.
For evaluating the tunneling current, we will only need $G_{{\bf
r}\,{\bf r'}}(\tau)$ for ${\bf r} = {\bf r'}$, which can
be expressed in terms of
the composite fermion Green's function as:
  \begin{equation}
 G_{{\bf r}\, {\bf r}}(\tau, \, a_{\mu}) =
e^{-i\int_{0}^{\tau}\,a_{0}({\bf r}, \, t)\, dt }
G^{CF}_{{\bf r} \, {\bf r}}(\tau, \, a_{\mu})\ .
\label{CS-transformation}
  \end{equation}
The exponential factor represents the action associated with
an extra flux $p\Phi_0$, which must be added at $t=0$ and removed at
$t=\tau$, to account for the difference between the creation operators for
an electron and a composite fermion.
Note that this factor makes the electron Green's function gauge
invariant, while the composite fermion Green's function itself
is not gauge invariant.
 
We employ a semiclassical approximation for
$G^{CF}_{\bf r\,r}(\tau,\,a_{\mu})$. To motivate it, think of an injected
electron which rapidly binds $p$ flux quanta and turns into a
composite fermion. The latter moves in the gauge field $a_{\mu}$
and picks the phase
  \begin{equation}
\label{phase}
\phi[a_{\mu}] = \int a^{\mu}({\bf r},\, t)\, j^{(CF)}_{\mu}({\bf r}, \, t)\,
d^2{\bf r}\, dt\ ,
  \end{equation}
where $j^{(CF)}_{\mu}({\bf r}, \, t)$ is the current describing
spreading of {\it free} composite fermion density.
Semiclassically in $a_{\mu}({\bf r},\, t)$ one writes
  \begin{equation}
\label{approximation}
G^{CF}_{{\bf r}\, {\bf r}}(\tau, \, a_{\mu}) =
e^{i\phi[a_{\mu}]}\, G^{0}(\tau)\ ,
  \end{equation}
where $G^{0}(\tau)$ is the composite fermion Green's function in
the absence of the slow gauge field. Note that fast fluctuations of
$a_\mu$ are included in $G^{0}(\tau)$ through renormalization of
Fermi-liquid parameters.  Below we average
the expression (\ref{approximation}) over the slow gauge field
(\ref{start}) and find $G_{\bf r\,r}(\tau)=\tau^{-\alpha}$.
 
Let us discuss the physical meaning of the phase factor
approximation (\ref{approximation}). It accounts correctly for
the action due to charge and gauge field relaxation during
tunneling. One notes the similarity to the classic infrared
catastrophe in Quantum Electrodynamics, where it
has been shown that the phase approximation gives correct
infrared asymptotics of electron Green's function\cite{Gorkov}.
Also, similar ideas were used in the problems of tunneling
coupled to an electromagnetic environment\cite{Nazarov}.
 
The most important correction to (\ref{approximation}) is due to
the shakeup of the composite Fermi system
due to short range interactions with  the
injected electron. This "orthogonality
catastrophe effect"\cite{Mahan} can be estimated by introducing
scattering channels for a given field $a_{\mu}({\bf r},\, t)$.
(For example, for the incompressible QH states the channels are
edge modes.) If the gauge field is concentrated on the scale
$L$, it excites $N\simeq k_FL$ composite fermion channels. In
terms of scattering phases $\delta_i$, ($i=1,...,N$), the
orthogonality contribution is given by $\exp(-\sum_{i}
\delta_i^2/\pi^2\, \ln(t/t_0))$, where $t_0$ is a short time
cutoff. To estimate $\delta_i$ we use Friedel sum
rule\cite{Mahan}: $\sum_i \delta_i = \pi$ (it means that we
inject unit charge). Hence, $\delta_i \sim \pi/N$ and the
orthogonality correction to the exponent $\alpha$ is $\simeq{\rm
const}/N\ll 1$.
 
The effective number of channels $N$ involved in charge
relaxation is certainly large in any compressible state, i.e.,
at finite $\sigma_{xx}$. However, the situation is more subtle
in an incompressible state with small number of edge modes,
where there are two possibilities. First, if all edge modes
propagate in the same direction, the orthogonality catastrophe
effect is simply absent, because in order to pick up the phase
$\delta_i$ a particle must travel {\it across} the perturbation.
This is the case, for instance, at Jain fractions $\nu=n/(2n+1)$
with $n>0$, where our result is exact. On the other hand, if
there is a small number of edge modes going in both directions,
the orthogonality correction is significant and may lead to
large deviations of $\alpha$ from the universal value. In this
way we can understand the non-universal values of $\alpha$
obtained for Jain fractions with negative $n$ in the absence of
impurity scattering\cite{Wen,Fisher}.
 
\noindent {\it Averaging over slow modes:\hskip2mm}
Putting together
Eqs.~(\ref{CS-transformation})-(\ref{approximation}) and
Eq.~(\ref{start}) we get
  \begin{equation} \label{G-j}
G_{\bf r\,r}(\tau)= G^0(\tau)\, \frac{\displaystyle \int Da_{\mu} \,
e^{-S_{eff}[a_\mu] + i\int j_{\mu}\,a^{\mu} \, d^2x\,dt }}
{\displaystyle  \int Da_{\mu} \, e^{-S_{eff}[a_\mu] } }
  \end{equation}
where the current  $j_{\mu}=  j^{(CF)}_{\mu}+  j^{(flux)}_{\mu}$, according to
Eq.(\ref{CS-transformation}),
includes the gauge field flux term
  \begin{equation}
j^{(flux)}_{\mu}({\bf r}, \, t)=-\delta_{\mu 0}
 \delta({\bf r})(\theta(t) - \theta(t-\tau))\ .
  \end{equation}
By expanding $S_{eff}[a_{\mu}]$ up to the second order in $a_{\mu}$ and
averaging over Gaussian fluctuations in (\ref{G-j}), we get
$G_{\bf r\,r}(\tau) = G^0(\tau) \exp\left(- S(\tau)\right)$ where
  \begin{equation}
\label{main}
S = \frac{1}{2}
\int d^3x\, d^3x'\, j^{\mu}(x)\,
j^{\nu}(x')\,\,
{\cal D}_{\mu\nu}(x, x') \ .
  \end{equation}
Here
${\cal D}_{\mu\nu}(x,\, x') =
\langle a_{\mu}(x)\, a_{\nu}(x') \rangle$
is the correlator of gauge field fluctuations ($x = ({\bf r}, t)$).
One may interpret $S(\tau)$ as the semiclassical action of spreading
charge\cite{LevitovShytov}.
 
\noindent {\it Calculation:\hskip2mm}
Hereafter we concentrate  on  the  case  of  local  conductivity
tensor $\hat{\sigma}$, i.e., assume $kl,\ \omega l/v_F<1$, where
$l$  is  the  mean  free  path  of  composite fermions. Then the
current $j_\mu^{(CF)}$ can be found from the diffusion  equation
with a particle source at ${\bf r}=0$\cite{Latin-Greek}
  \begin{equation}
\label{diffusion}
\partial_t j_0 + \nabla_{i}(D_{ij}\nabla_{j} j_0)  =
\delta^{(2)}({\bf r})\bigl(\delta(t) - \delta(t-\tau)\bigr)\ ,
  \end{equation}
Here  the  {\it  noninteracting}  composite  fermion  diffusion
tensor ${\hat D}$ is given by the Einstein relation: $\hat D=\kappa_0
\hat{\sigma}^{(0)}$. The current is determined by  the  diffusion
law ${\bf j} = -\hat{D}\,\nabla j_0$.
 
To get the electron Green's function at the QH state edge,
we choose the $x$-axis along the edge and assume that
tunneling occurs at $x = 0$, $y = 0$ (2DEG occupies the half-space $y>0$).
Then Eq.(\ref{diffusion}) must be solved with the boundary condition
$j_y(y=0) = 0$. To treat Eq.(\ref{diffusion}),
it is convenient to use Fourier transform with respect
to $x$ and $t$: $j_0(k, y,\omega) = \int
j_0(x,y,t)\, e^{ikx - i \omega t} dxdt$.
One can write a formal solution to Eq.(\ref{diffusion}):
  \begin{equation}
\label{formal}
j_0 = \frac{1}{i\omega + \nabla_i D_{ij}\nabla_{j}}\, J(\omega)\delta(y)
\ , \ J(\omega) =( 1-e^{-i\omega\tau})
  \end{equation}
Eq.(\ref{formal}) holds also in the general case if $\hat{D}$ is
treated as a nonlocal operator.
 
To  find  the  gauge  field  correlator  ${\cal D}_{\mu\nu}$, we
choose the gauge $a_0 = 0$ and, following \cite{HLR}, write
  \begin{equation}
\label{D-function}
\hat{{\cal D}} = \left(\hat{K}_0 + \hat{U}^{-1}\right)^{-1}\ .
  \end{equation}
Here
  \begin{equation}
\hat{U}_{ij} =  \nabla_i\,\frac{U({\bf r} - {\bf r}')}{\omega^2}\, \nabla_j' +
\frac{2\pi\hbar i p}{e^2\omega}\,\epsilon_{ij}\,\delta({\bf r}-{\bf r'})
  \end{equation}
is the bare interaction
and $\hat{K}_0$ is current correlation function which can be found, e.g.,
using fluctuation-dissipation theorem:
$
\hat{K}_{0} = \omega^2(i\omega + (\nabla\hat{D})\nabla)^{-1}\hat{\sigma}_{0}
\ .
$
For  simplicity,  we  assume  short-range  electron  interaction
$U({\bf r})= U\delta({\bf r})$.
 
Let us remark that one has to be careful in
substituting (\ref{D-function}) into (\ref{main}),
since the flux current does not
couple to the Coulomb part of the gauge field.
Hence, one must subtract a flux term from the
longitudinal current: $\nabla \cdot {\bf j} \to \nabla \cdot {\bf j} -
\partial_t j^{(flux)}_0$.
 
After that, we decompose the product of two operators into partial fractions:
  \begin{eqnarray}
\label{S0}
S(\tau)  =  \frac{1}{2} \Biggl\langle\, J\, \Biggl|
\big(\nabla(\hat{\sigma}\nabla)\big)^{-1}
\Bigl(
\frac{1}{i\omega + (\nabla\hat{D})\nabla} -
\\
\nonumber
\frac{1}{i\omega + (\nabla\hat{D})\nabla + \omega^2
\hat{\sigma}_0\hat{U}}
\Bigr)\Biggr|\, J\,\Biggr\rangle
\end{eqnarray}
(angular brackets denote integration over coordinates.)
Now, the operators in (\ref{S0}) must be inverted, taking into account
the boundary condition
$j_{y}(y=0) = 0$. After some tedious algebra one finally arrives at
  \begin{equation}
S(\tau) = \sum\limits_{\omega, k} \frac{|J(\omega)|^2}{2\omega}
\left(\frac{1}{\sigma_{xx} q + i\sigma_{xy}k} -
\frac{1}{\sigma_{xx}^{(0)}\, q_{0} + i\sigma_{xy}^{(0)} k}\right)\ ,
\label{action}
  \end{equation}
where $\sigma_{xx}$ and $\sigma_{xy}$ are components of the measured
conductivity tensor\cite{HLR}
$\hat{\sigma}=\hat{\rho}^{-1}$,
while $q^2 = k^2 + i\kappa\omega/\sigma_{xx}$,
and $q_{0}^2 = k^2 + i\kappa_0\omega/\sigma_{xx}^{(0)}$.
Integrating (\ref{action}) over $\omega$ and $k$ one gets
$S(\tau) = (\alpha-1)\ln(\tau/\tau_0)$ with $\alpha$ defined by
Eq. (\ref{Alpha}). Here $\tau_0$ is a short-time cutoff.
 
We also assume that composite fermions are well defined quasiparticles
and thus $G_{CF}^0(\tau) \sim \tau^{-1}$.
To justify that, one may consider the self-energy
$\Sigma(\epsilon, \xi)$ of a composite fermion arising from the fluctuations
of the gauge field. The self-energy is known to be a singular
function of energy $\epsilon$ but not of $\xi = v_F(|{\bf p}| - p_F)$ (see \cite{HLR},
\cite{effective-mass}).
Because of that, the pole of the equal point Green's function
$G^{0}_{{\bf rr}}(\tau)$ remains intact. This situation is similar
to that for electron-phonon
interaction, where $\Sigma$ does not depend on $\xi$ and thus
does not affect tunneling current.
We remark that the point ${\bf r}$ is taken to be close to  the
edge of the system, and the form of $G^{0}$ holds in this case even for
$\delta B \ne 0$.

Putting    everything    together,   we   have   $G(\tau)   \sim
\tau^{-\alpha}$.  The  tunneling  current  is  then   given   by
\cite{tunneling-current}:
  \begin{equation}
I(V) \sim
{\rm Im} \int\limits_{0}^{\infty} G(\tau)\, \frac{e^{ieV\tau}}{\tau}\, d\tau
\sim V^{\alpha}
\label{current}
  \end{equation}
where $V$ is the voltage across the barrier.
 
Now we check that the number of channels is big. The gauge field
disturbance  caused by the tunneling electron decays in the bulk
as $\exp(-qy)$. Hence, after time $t$ the  perturbation  spreads
over  $y_0(t)  \simeq\sqrt{\kappa\sigma_{xx}  t}  $. To estimate
spreading  along  the  edge,  one  compares  two  terms  in  the
denominator  of  (\ref{action}) assuming $k\sim x_0^{-1}$, which
gives $x_0(t) \simeq\sigma_{xy}/\sigma_{xx} y_0(t)$.  (Note  that
charge  spreading is effectively one-dimensional if $\sigma_{xy}
\gg \sigma_{xx}$.) To estimate the number of scattering channels
parallel to the boundary, one  takes  $L  \sim  y_0(t)$  as  the
spatial  scale,  and  gets $N \sim y_0(t)k_F$. Since $y_0(t)$ is
large at large $t$ (i.e.,  at  sufficiently  low  bias  $V  \sim
\hbar/t $) we indeed have $N\gg 1$.
 
\noindent {\it Conclusion:\hskip2mm}
  The above composite fermion theory of tunneling into a Quantum
Hall edge treats on a similar footing both compressible and
incompressible states. We have shown that tunneling $I-V$
characteristic is non-ohmic and obeys a power law for both
compressible and incompressible Quantum Hall edges. The power
law exponent is expressed in terms of $\rho_{xx}$ and
$\rho_{xy}$, with a weak dependence on the ratio $\kappa /
\kappa_0$, which depends on the form of the interaction. The
exponent is a continuous function of $\rho_{xy}$, with cusp-like
sigularities at $\nu=1/2$ and $1/4$ in the limit where
$\rho_{xx}=0$. The physical origin of the singularity is that
the effective magnetic field seen by composite fermions changes
sign at $\nu=1/2$ and $1/4$.
 
It is shown that the model is robust for compressible QH states
when the number of scattering channels involved in charge
relaxation is large. At the Jain fractions $\nu=n/(2n\pm 1)$ the
results agree with the chiral Luttinger liquid theories.
 
The calculated exponent is a monotonically decreasing function
of filling fraction, in qualitative agreement with the recent
experiment by Chang et al\cite{Chang2,Chang3}. However, the
theoretical value of $\alpha$ is larger than that observed in
the experiments, particularly near $\nu=1/2$. The discrepancy
might be explained if the density near the edge of the 2D
electron system is actually somewhat larger than that in the
bulk.

\acknowledgements
\noindent
This work was stimulated by discussions with Albert Chang. We
thank Bell Labs and the NSF for support (grant DMR94-16190 and
the MRSEC Program under award  DMR94-00334).

%\figure{\label{fig1}
%{\it  Fig. 1 }
%The tunneling exponent (\protect{\ref{Alpha}}) is shown as a
%function of $\rho_{xy}$ with the composite fermion flux $p=2$
%for $1\le\rho_{xy}\le 3$ and $p=4$ for $3\le\rho_{xy}\le5$.
%Constant Hall angle is assumed. For $\rho_{xx}=0.05\rho_{xy}$
%the exponent is plotted for three values of the model
%short-range interaction $U=1/\kappa-1/\kappa_0$. At $\rho_{xx}=0$
%the exponent is universal (no $U-$dependence), but at finite
%$\rho_{xx}$ it can be either bigger or smaller than the
%universal result, depending on the interaction strength
%$\kappa_0U$.
%   }

%\input epsfig.sty
%\input epsf.sty
%\begin{center}
%\begin{figure}
%\figure
%\label{fig1}
%\epsfig{file= figure1.eps, height= 130mm,  angle= 270}
%\epsffile{figure1.eps}
%\caption{
%The tunneling exponent (\protect{\ref{Alpha}}) is shown as a
%function of $\rho_{xy}$ with the composite fermion flux $p=2$
%for $1\le\rho_{xy}\le 3$ and $p=4$ for $3\le\rho_{xy}\le5$.
%Constant Hall angle is assumed. For $\rho_{xx}=0.05\rho_{xy}$
%the exponent is plotted for three values of the model
%short-range interaction $U=1/\kappa-1/\kappa_0$. At $\rho_{xx}=0$
%the exponent is universal (no $U-$dependence), but at finite
%$\rho_{xx}$ it can be either bigger or smaller than the
%universal result, depending on the interaction strength
%$\kappa_0U$.
%}
%\end{figure} 
 

\begin{references}
\bibitem{Latin-Greek}
We use Greek letters for space-time components and Latin  for
the  space only.
\end{references}
\end{document}